\documentclass{llncs}

\usepackage{graphicx}
\usepackage{url}
\usepackage{paralist}
\usepackage{relsize}
\begin{document}

\title{Domain-Specific Modeling and Code Generation for Cross-Platform Multi-Device Mobile Apps\thanks{This research, developed under the supervision of Professor Marco Brambilla, is on initial phase.}}
\author{Eric Umuhoza}
\authorrunning{Eric Umuhoza}   
\institute{
Politecnico di Milano. 
Dipartimento di Elettronica, Informazione e Bioingegneria \\
Piazza L. Da Vinci 32.
I-20133 Milan, Italy\\
\email{eric.umuhoza@polimi.it}
}

\maketitle

\begin{abstract}
Nowadays, mobile devices constitute the most common computing device. This new computing model has brought intense competition among hardware and software providers who are continuously introducing increasingly powerful mobile devices and innovative OSs into the market.
In consequence, cross-platform and multi-device development has become a priority for software companies that want to reach the widest possible audience. However, developing an application for several platforms implies high costs and technical complexity. Currently, there are several frameworks that allow cross-platform application development. However, these approaches still require manual programming. My research proposes to face the challenge of the mobile revolution by exploiting abstraction, modeling and code generation, in the spirit of the modern paradigm of Model Driven Engineering.
\end{abstract}

\section{Introduction}
\label{sec:intro}
\textbf{Context}
Nowadays, mobile devices constitute the most common computing device.
A vast array of features has been incorporated into those devices to address the different demands of users spanning from games to serious business.
Today mobile devices are as powerful as desktop computers in terms of their computing capabilities.
This new computing model have brought intense competition and innovation among devices, OSs, and application providers.\\\\
\textbf{Problem}
Even though the mobile OS market is beginning to mature and consolidate, most researches concur that it is unlikely that a single vendor will dominate the future mobile-centric world~\cite{HeitkotterHM12a}.
The dilemma between browser-based (HTML 5) and native (iOS, Android, Blackberry, Symbian, and Windows Phone) interfaces remains relevant and will challenge the capacity of organizations to meet the increasing demand for mobile apps.

Moreover, the vastness and diversity of mobile devices and operating systems available on the market oblige companies to produce and deploy the same app several times, once for each of the different mobile platforms.
Unfortunately, cross-platform and multi-device development is a barrier for today's IT solution providers, especially SMEs, due to the high cost and technical complexity of targeting development to a wide spectrum of devices, which differ in format, interaction paradigm, and software architecture.

Currently, there are several frameworks implementing different methodologies for cross platform application development (Web, Hybrid, Interpreted and Cross Compiled): examples include PhoneGap(Cordova)  
\footnote{www.phonegap.com}, Appcelerator Titanium\footnote{www.appcelerator.com}, and Xamarin\footnote{www.xamarin.com}.
Unfortunately, these approaches still require manual programming which yields to \textit{high risks of errors}, \textit{inconsistencies} and \textit{inefficiencies}.\\\\
\textbf{Relevance}
The number of apps that are available in the online markets has reached unseen numbers. In fact, by July 2014, the Google Play store counted 1.3 million of available apps while Apple's App Store counted 1.2 million. In parallel with these numbers, the market also expects an increase in the number of global smart-phones users, which is expected to surpass 2 billion by 2016 ~\cite{emarketer} in comparison with 1.4 billion users estimated in 2013 ~\cite{statista}. From those number we can expect a healthy market of software apps that would be powered by a steady increase in the number of mobile device users, which as of today have, on average, 41 apps installed on their devices ~\cite{flurry}. 
Furthermore, the motivation of the software development companies to continue producing more and better apps is supported by recent industry figures, according to which global mobile app revenues are projected to surpass 76.52 billion U.S. dollars in 2017. 
ABI research forecasts in 2018, app revenues will be worth 92 billion  U.S dollars~\cite{ABI}.\\\\ 
\textbf{My Vision}
My research proposes to face the challenge of the mobile revolution by exploiting \textbf{abstraction}, \textbf{modeling} and \textbf{code generation}. Different MDD approaches for cross-platform mobile apps development will be studied with the aim of providing a framework allowing apps developers to choose a MDD approach that meets their requirements. The problem of multi-device will be approached by providing a set model-to-model transformations that are applied over the same model of the mobile app. Each of these transformations will produce a new model that
describes the shape the app will have in a particular device family (tablet, smart-phone, smart-watch, etc).

The paper is organized as follows: Section \ref{sec:problem} describes the problem that my research intends to solve; Section \ref{sec:relwork} reviews the related work; \ref{sec:approach} presents the proposed solution; Section \ref{sec:approach} presents the methodology and tools that will help to use the proposed solution; and Section \ref{sec:status} presents the preliminary work, the future works and the contributions expected from my research.

\section{Problem Statement}
\label{sec:problem}
Cross-platform and multi-device development is a barrier for today's IT solution providers, especially SMEs, due to the high cost and technical complexity of targeting development to a wide spectrum of devices, which differ in format, interaction paradigm, and software architecture. The challenges of mobile apps developments that my research attempt to address are described in detail in the next paragraphs.
\\\\
\textbf{P1: Platform.}
The market of mobile operating systems is fragmented and rapidly changing.
The diversity of OSs available on the market oblige software developers that want to reach a large audience of users to develop their apps for each platform (at least for the most competitive ones such as Android, iOS, and Windows Phone~\cite{gartner2012,HeitkotterHM12a}) separately.\\\\
\textbf{P2: Different Front End Requirements.}
From the user interactions perspective, mobile apps are expected to support a wider set of interactions that are captured by means of a tactile surface (\textit{interaction through a set of gestures like taps and swipes}) and through the different sensors that are packed into the device (\textit{sensor-based interactions like rotate and shake}).
Moreover, front-end design of mobile apps must consider the size constraints imposed by the characteristics of the screens of modern mobile devices. In addition, the mobile apps must adapt to changes of the context (the communication network, the battery level of the device and the environment surrounding the user) to deliver the most efficient interface~\cite{BrambillaMU14}.\\\\
\textbf{P3: Resource Scarcity.}
Even though the mobile OSs provide the optimization strategies to cope with the scarcity of resources on mobile devices (like  memory and storage, battery, and the instability and diversity of the communication networks), mobile apps need to be able to receive system notifications (such as the battery level and new networks availability) and react appropriately to them in order to provide a consistent and reliable user experience.  
\\\\
\textbf{P4: Device Diversity.}
Currently, the market offers several families of mobile devices such as tablets, smart-phones and emerging smart-watches. Each of them can show different amounts of information, uses particular navigation patterns and has a diverse set of sensors available at run-time. These differences imply that software developers have to create applications that can either adapt to the specific device in which they are running, or create different versions of the application, each of them targeting a specific device family. 

\section{Related work}
\label{sec:relwork}
\vspace{-5pt}
This research is the first one that attempts to compare various model-driven
strategies for cross-platform and multi-device mobile apps development with aim of providing guidelines to the developers who need to adopt MDD approach in their mobile apps development process. Thus, this section assesses the existing works that apply the MDD approach to the development of mobile apps in a broad sense. 

Those works can be divided into two different clusters. 
On one hand we encounter a corpus of research that apply model-driven techniques to specify application interfaces and user
interaction (in a broad sense) for multi-device UI modeling.
Among them we can cite: \begin{inparaenum}[]
\item TERESA(Transformation Environment for inteRactivE Systems representations)~\cite{BertiCMPS04}, based on a so-called \emph{One Model, Many Interfaces} approach to support model based GUI development for multiple devices from the same ConcurTaskTree (CTT) model; 
\item  MARIA~\cite{PaternoSS09}, another approach based on CTT;
\item UsiXML (USer Interface eXtended Markup Language)~\cite{Vanderdonckt2005};
\item  Unified Communication Platform (UCP); and 
\item IFML (Interaction Flow Modeling Language)~\cite{ifml}, a platform independent modeling language designed to express the content, user interaction, and control behavior of the front-end of software applications. However, none of them specifically addresses the needs of mobile apps development. 
\end{inparaenum}

My research will leverage on the IFML language which has been recently adopted as a OMG standard. In particular my research provided a mobile extension~\cite{BrambillaMU14} of IFM that will be used to describe, at PIM level, the main aspects of a mobile app front-end. Moreover, since IFML can be used in tandem with other modeling languages, aspects like the domain model and the business logic of the app will be  defined through standard languages like UML.

On the other hand we find a collection of works that proposes MDD solutions for the development of cross-platform  mobile apps. This cluster of researches can be further divided into three groups depending on whether they  produce native, hybrid or web-based apps.
 
In the first group we encounter projects like MD2~\cite{HeitkotterMK13}, an approach that focuses on the code generation (for Android and iOS) of data-driven business apps for tablets according to the MVC paradigm, Vaupel et al.~\cite{vaupel2014model} defined an infrastructure that supports the specification of different variants of an Android app according to user roles, and Franzago et al.~\cite{FranzagoMM14} defined a collaborative framework for the design and development of data-intensive mobile apps. Their approach is based on PIM languages allowing the specification of various viewpoints (navigation, content, user interface, and business logic) of a data-intensive mobile apps. In contrast with MD2, which only targets Android and iOS, my research will allow for the generation of hybrid and native apps. Moreover, MD2 offers a textual syntax that is suitable for users with a programming background. My research instead will leverage on the graphic syntax provided by IFML to enable domain experts participate in the design and specification of mobile apps. Finally, my research is different from Franzago et al. because it uses standard languages like IFML and UML for modeling the different concerns of mobile apps. Moreover my proposed solution uses a single modeling language to specify the UI and navigation, whereas Franzago et al. use one language for each concern. 
 
The second group concerns solutions that generate hybrid apps to address the cross-platform issue. As part of this group we can mention Applause\footnote{\url{ https://github.com/applause/applause}}, a domain-specific language and a set of code-generators to produce mobile apps for iPhone, Android, Windows Phone on top of Google App Engine. As mentioned before, my solution will consider the generation of both  native and hybrid apps and will provide a graphical syntax, which are both missing in Applause. 
 
Finally, the third group contains MDD proposals that generate web-based mobile apps. A good representative of this category is Mobl\footnote{\url{http://www.mobl-lang.org/}}, an open source language designed to speed up building mobile apps. Mobl offers a concise language to build native-feeling web apps for mobile system. Even though this approach towards cross-platform development is not considered in my solution, it could be seen as a complementary strategy suitable for specific scenarios where the native or the hybrid solutions do not bring any additional value.

\section{Proposed Solution}
\label{sec:approach}
I propose to face the challenge of cross-platform multi-device mobile apps development by applying a model driven development (MDD) approach.
I will use the model driven architecture (MDA)\footnote{\url{//http://www.omg.org/mda/}} as a reference framework to illustrate the proposed solution.
MDA defines models at three different levels of abstraction: Computation Independent Models (CIM), Platform Independent Models (PIM), and Platform Specific Models (PSM). A set of mappings between each level and the subsequent one can be defined through model transformations. Every CIM can map to different PIMs, which in turn can map to different PSMs but many other combinations can be followed, for instance skipping one of the levels.

My research proposes to address the problem of cross-platform mobile apps development through four code generation alternatives depicted in Figure \ref{fig:crossplat}.  The code generators input the models that describe the app requirements and output the app code. The domain model and the business logic of the app are defined through UML while the app front-end is defined through IFML.
In contrast I will deal with the issue of  multi-device mobile apps development through model transformations that would be applied at the PIM level.
Section \ref{cross} is dedicated to the issue of cross-platform mobile apps development (\textit{P1, P2 and P3}) while the solution to the problem of multi-device development (\textit{P4}) will be presented separately in Section \ref{device}.
\subsection{Cross-Platform Development}
\label{cross}
When following the MDD approach, several code generation strategies are possible depending, both, on the abstraction level to be used when modeling the application and the abstraction level of the code to be generated. 

\begin{figure}[h]
\begin{center}
 \includegraphics[width=.75\textwidth]{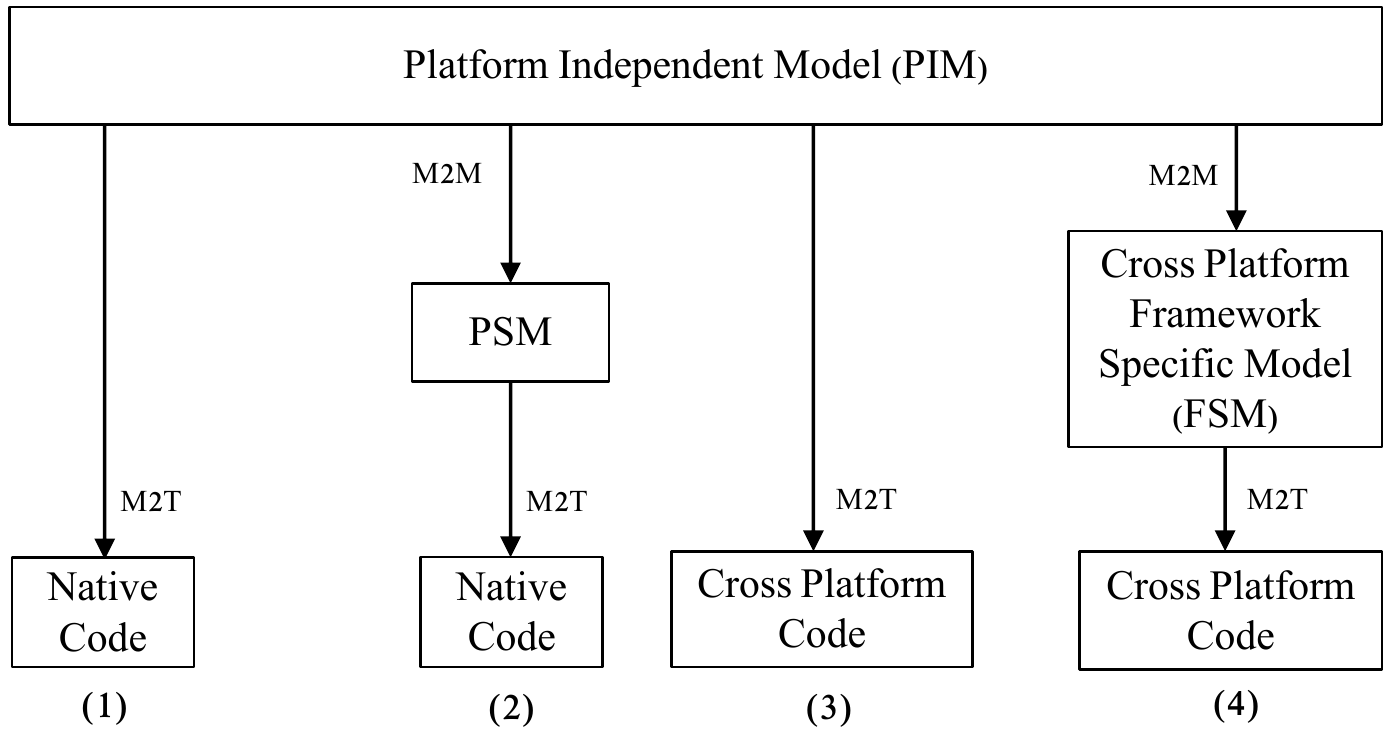}
 \vspace{-5pt}
 \caption{MDD approaches for cross-platform mobile apps development: the requirements of the app are described in a PIM from which the app code is generated through the application of a set of model-to-model(M2M) and/or model-to-text (M2T) rules.}
\label{fig:crossplat}
\end{center}
 \vspace{-10pt}
\end{figure}

In this research, I will analyze all four alternatives for cross-platform mobile apps development (figure \ref{fig:crossplat}) with the aim of providing apps developers the guidelines to choose the right MDD approach for them according to their particular requirements. To reach this goal, the pros and cons of each of those options will be studied.

\begin{enumerate}[(1)]
\item {PIM-to-Native Code}.
By following this option, the app requirements shall be specified through a Platform Independent Modeling Language such as mobile IFML ~\cite{BrambillaMU14}. Using a Platform Independent Modeling Language shall allow the modeling of the specific Front End requirements that characterize mobile applications (\textbf{P2}), as well as the interactions that occur between the application and the Operating System (\textbf{P3}). The last aspect is important because these type of interactions allow mobile apps to react to changes in their execution environment. Finally the creation of a cross-platform (\textbf{P1}) app is achieved by providing different native code generators, one for each targeted platform, that take as input the same PIM describing the app and generate as outputs the code for the corresponding platforms. For example to create a native iOS and Android application using this approach will require the implementation of two code generators that are able to produce Objective-C and Java code from the same model of the application.

\item {PIM-to-PSM-to-Native Code}.
Like in the previous option the front end requirements of the application is defined through a Platfrom Independent Model. In this case, however, the PIM is first transformed into different PSMs, each of which refines the initial model adding the platform specific details that are not captured at the PIM level. Once the PSMs have been produced, a set of simple code generators transforms these models into the native code of each of the target platforms. The previous means that besides a set of code generators that produces the native code of each of the target platforms, a Platform Specific Modeling Language for each of them will also be needed. Despite of this added cost, the introduction of an additional PSM level addresses the problems of platform diversity (\textbf{P1}), front-end requirements (\textbf{P2}) and resource scarcity (\textbf{P3}).

\item {PIM-to-Cross Platform Code}.
By following this option the application requirements are specified in a Platform Independent Model from which the code is generated. In this case, however, the generated code must conform to the structure of a particular \emph{cross-platform framework}. Then, it will be the responsibility of the framework to guarantee that the generated app will run across the different platforms. To achieve this the framework will typically take the generated code and produce the binary files for each of the target platforms using an automated process.

This option simplifies the creation of a cross-platform app (\textbf{P1}) because it generates the code for a cross-platform solution. For example to create an app that will run both on Android and iOS with \textit{PhoneGap} as the cross-platform framework a software designer will start by creating the Platform Independent Model of the application. Then, using a single M2T transformation he will produce the \textit{HTML5, CSS and Javascript} code required by  PhoneGap. Finally, using the build tools offered by the framework the code will be transformed into the binary files required by each of the target platforms, which in our case are the ipa file for iOS and apk file for Android.

A Platform Independent Modeling language shall allow the modeling of the specific Front End requirements that characterize mobile applications (\textbf{P2}), as well as the interactions that occur between the app and the OS (\textbf{P3}).

\item {PIM-to-Framework Specific Model (FSM)-to-CPC}.
With respect to the previous option, this approach introduces the FSM which gathers the information regarding the \emph{cross platform framework} (such as PhoneGap, AppCelerator Titanium, and Xamarin) used to produce the apps. FSM is a PSM in which the \emph{Platform} in the MDA terminology, is actually a \textit{cross-platform framework} for mobile apps development.
In this case the PIM is first transformed into the FSM which refines the initial model adding the \textit{cross-platform framework} specific details that are not captured at the PIM level. Once the FSM has been produced, a simple code generator transform that model into the code required by the \textit{cross-platform framework}. The introduction of FSM level requires to provide a FSM modeling language.
Similarly to the previous option, the problem of front-end requirements (\textbf{P2}) and resource scarcity (\textbf{P3}) are addressed at modeling level while the problem of platforms diversity is achieved by the \textit{cross-platform framework} itself. 
\end{enumerate}

\begin{figure}[h]
\begin{center}
 \includegraphics[width=.7\textwidth]{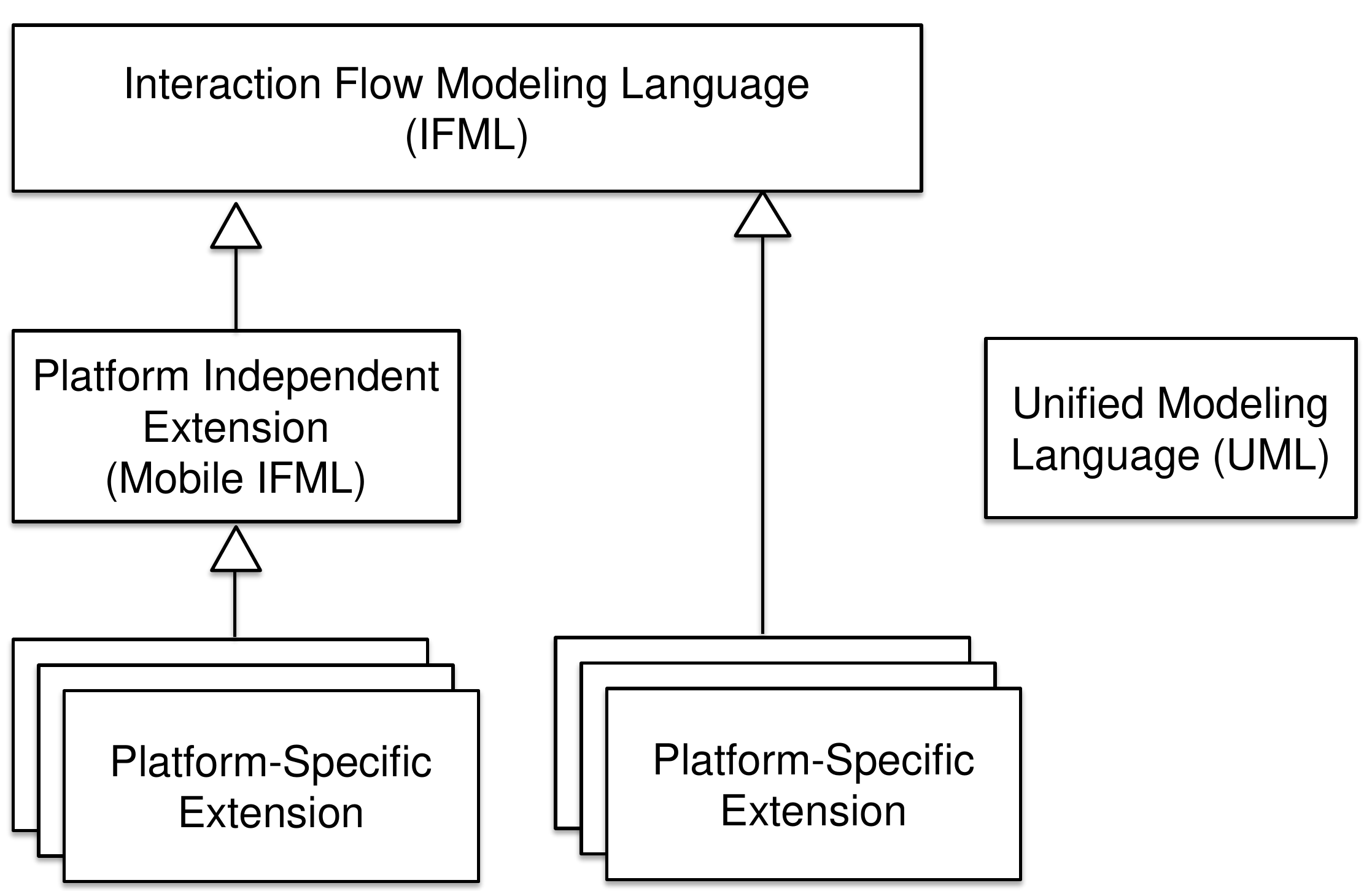}
 \vspace{-10pt}
 \caption{Modeling languages stack: mobile IFML defined as a mobile extension of IFML shall be used to model the front-end of the app at the PIM level. The required PSM languages will be defined as extensions of the mobile IFML. While the domain model and application logic will be specified through UML}
\label{fig:lang}
\end{center}
 \vspace{-10pt}
\end{figure}

All the options mentioned so far require a modeling language allowing the specification of application requirements in a platform independent manner. In this research I will use an OMG standard, the Interaction Flow Modeling Language (IFML) as a reference PIM language. In particular this research will define a mobile extension of IFML to allow the modeling of the specific Front End requirements that characterize mobile applications, as well as the interactions that occur between the application and the OS. The PSM languages required by options \textit{(2)} and \textit{(4)} will be defined as Platform Specific Extensions of the mobile  extension of IFML mentioned before. To clarify the solution, Figure  \ref{fig:lang} shows a diagram with the proposed modeling stack.

\subsection{Multi-Device Development}
\label{device}
\vspace{-3pt}
The solution I propose to address the issue of device diversity (\textbf{P4}) is based on the following assumption: the characteristics of different devices of the same family (tablet, smart-phone, smart-watch, etc) do not change drastically. For instance different smart-watches are assumed to have more or less the same screen dimensions and different smart-phones are assumed to have have roughly the same sensors.  

\begin{figure}[t]
\begin{center}
 \includegraphics[width=.55\textwidth]{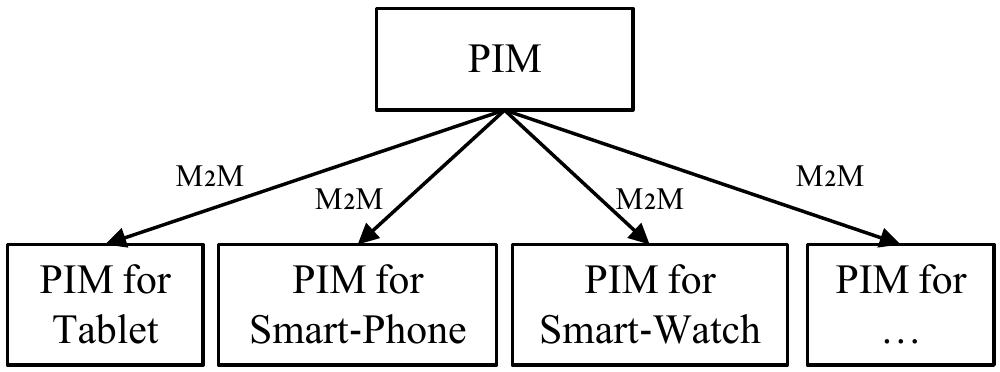}
  \vspace{-5pt}
 \caption{Multi device development: The app requirements are described in a \textbf{general} PIM. Different model transformations are applied to that PIM to produce a new PIM that describes the shape of the app in each of the target device families.  The M2M transformations can be conceptualized for instance through a set of ATL \cite{ATL} rules}
\label{fig:multdev}
\end{center}
 \vspace{-10pt}
\end{figure}

To deal with this issue, I propose a strategy based on a set of model to model transformations (M2M) that are applied over the same general model of the mobile application. Each of these transformations will produce a new model that describes the shape the application will have in a particular device family (Figure \ref{fig:multdev}). For example, when designing an application that should be used in a tablet, a smart-phone and a smart-watch, a software designer will first create a single \textit{general} model of the application. Then, he will use a particular M2M transformation to generate a version of the model that is suitable for phones. He will then repeat the same process using different M2M transformations to obtain the model for the tablet and the watch. At the end of the day, the software designer will have four different Platform Independent Models, that can be transformed into running code following any of the previously discussed strategies.

It is important to highlight that the aforementioned M2M transformations could be defined using two different approaches. The first approach is based on the definition of a fixed set of transformation rules that given a \textit{general} model, are able to produce the model suitable for a particular device family. In the second approach, the M2M transformation rules are defined by the software designer with the support of a model editor that records at each step the modifications he applies over the general model to obtain the model of a particular device family~\cite{sun2009model,wimmer2007towards,varro2006model}. In summary, in the first approach the transformation rules are fixed and application independent whereas in the second approach the transformation rules are defined by the software designer and depend on the application. During my research I will evaluate which of the two strategies yields the best results.

\section{Methodology and Tools}
\label{sec:methodology}
The results of a comparative study among different MDD approaches to cross-platform and multi-device mobile apps development (Section \ref{sec:approach}) will constitute guidelines for apps developers to choose a MDD approach that fits their requirements.

Front-end design of mobile apps is a complex task, the content and the navigation among them must be well designed at the purpose of exploiting at best the limited space available.
Providing design patterns for both the content organization and navigation could help application designers to find solutions to common design challenges and to reuse them.
My research will provide a set of design patterns as to illustrate and simplify the modeling of mobile app using the defined method and tools.

\section{Status and Future Works}
\label{sec:status}
\subsubsection{Preliminary Work.}
My research starts from a deep state-of-the-art analysis on mobile apps development in a wide sense on model driven approaches in particular. The current implementations include: 
\begin{itemize}
\item A platform Independent Modeling language for mobile apps~\cite{BrambillaMU14} along with its graphical modeling tool\footnote{https://github.com/mobileIFML/ifml-editor}, an eclipse plugin based on Sirius;
\item A set of first prototypes of code generators both for native platforms and cross-platform frameworks;
\item Initial validation through developed mobile apps that includes Instangram, CamScanner, and Foursquare.
\end{itemize}
\textbf{The future works} include:\begin{itemize}
\item {Model-driven analytic}.
I will study how model-driven techniques can be combined with existing analytic in order to provide a rich message from web and mobile apps monitoring;
\item {Multi-Devices Development}.
The issue of model to model transformations (M2M) mentioned in Section 3.2 needs a deep investigation to understand whether is better to rely on application independent and fixed M2M rules, application dependent rules or a combination of both~\cite{czarnecki2003}.
\item {Platform-Specific Extensions}.
The mobile language defined, allows the modeling the app in a platform independent manner. However, in some cases (options \textit{(2)} and \textit{(4)} figure \ref{fig:crossplat}) it could be worth to model the app or some part of it taking into account some platform specific detail that are not captured at the PIM level;
\item {Mobile-Specific Design Patterns and Anti-Patterns}.
Identification and modeling of common design patterns for model-driven mobile applications design and identification of mobile-specific anti-patterns (common design patterns not suitable for mobile world);
\item {Modernization}. I will study how to transform legacy applications into mobile apps. The design patterns will be useful in this phase;
\item {Use Cases}. I will provide a portfolio of B2C and B2B vertical mobile apps, demonstrating the effectiveness of the research approach. 
\end{itemize}
\subsubsection{Expected Contributions.}
My research is expected to make three main contributions. Firstly, it will give a comprehensive overview of current MDD approaches for cross-platform app development. Secondly, it will provide a framework of criteria for evaluating MDD approaches to mobile apps development. The proposed criteria could be used for future assessments. Thirdly, it will provide decision advice allowing developers to choose a MDD approach that better fits their requirements. \\
In practice the contribution of my research can be appreciated from two different perspectives. First, from the point of view of the software development companies who could appreciate a MDD approach to cross-platform mobile applications development that can reduce the technical complexity and the development costs~\cite{Diaz20101970}. From the perspective of the final users, having access to cross-platform apps that are available in the different online markets will let them choose freely the type of device and OS they can use, without worrying about the availability of particular apps for their device.
\vspace{10pt}

\footnotesize
\bibliographystyle{plain}
\bibliography{codegenerator}

\end{document}